\documentclass[fleqn,usenatbib]{mnras}
\usepackage{newtxtext,newtxmath}
\usepackage[T1]{fontenc}

\DeclareRobustCommand{\VAN}[3]{#2}
\let\VANthebibliography\thebibliography
\def\thebibliography{\DeclareRobustCommand{\VAN}[3]{##3}\VANthebibliography}

\usepackage{orcidlink}
\usepackage{graphicx}
\usepackage{amsmath}
\usepackage{multirow}

\title[High-energy Neutrinos from AGN BBH mergers]{High-energy Neutrinos from Merging Stellar-mass Black Holes in Active Galactic Nuclei Accretion Disk}

\author[J.-P. Zhu]{
Jin-Ping Zhu\orcidlink{0000-0002-9195-4904}$^{1,2}$\thanks{E-mail: jin-ping.zhu@monash.edu}
\\
$^{1}$School of Physics and Astronomy, Monash University, Clayton Victoria 3800, Australia\\
$^{2}$OzGrav: The ARC Centre of Excellence for Gravitational Wave Discovery, Australia
}

\date{Accepted 2023 November 17. Received 2023 November 17; in original form 2023 September 29}
\pubyear{2023}

\begin{document}
\label{firstpage}
\pagerange{\pageref{firstpage}--\pageref{lastpage}}
\maketitle

\begin{abstract}

A population of binary stellar-mass black hole (BBH) mergers are believed to occur embedded in the accretion disk of active galactic nuclei (AGNs). In this {\em Letter}, we demonstrate that the jets from these BBH mergers can propagate collimatedly within the disk atmosphere along with a forward shock and a reverse shock forming at the jet head. Efficient proton acceleration by these shocks is usually expected before the breakout, leading to the production of TeV$-$PeV neutrinos through interactions between these protons and electron-radiating photons via photon-meson production. AGN BBH mergers occurring in the outer regions of the disk are more likely to produce more powerful neutrino bursts. Taking the host AGN properties of the potential GW190521 electromagnetic (EM) counterpart as an example, one expects $\gtrsim1$ neutrino events detectable by IceCube if the jet is on-axis and the radial location of the merger is $R\gtrsim10^5R_{\rm{g}}$, where $R_{\rm{g}}$ is the gravitational radius of the supermassive BH. Neutrino bursts from AGN BBH mergers could be detected by IceCube following the observation of gravitational waves (GWs), serving as precursor signals before the detection of EM breakout signals. AGN BBH mergers are potential target sources for future joint GW, neutrino, and EM multi-messenger observations.

\end{abstract}


\section{Introduction}

It has been proposed that a significant number of stars and compact objects, including white dwarfs, neutron stars (NSs), and BHs, can be harbored within the AGN accretion disks around the supermassive BHs (SMBHs). These AGN stars and compact objects can be either captured from the nuclear star clusters \citep[e.g.,][]{Syer1991,Artymowicz1993,Fabj2020,MacLeod2020} or by in situ formation \citep[e.g.,][]{Kolykhalov1980,Shlosman1989,Goodman2004,Wang2011,Wang2012,Dittmann2020,Fan2023}. The AGN disks offer a natural environment for embedded stars and compact objects to grow, accrete materials, and undergo orbital migration within disks \citep[e.g.,][]{Mckernan2012,Bellovary2016,Perna2021AccretionInduced,Jermyn2021,Wang2021,Dittmann2021,Pan2021,Tagawa2022,Kaaz2023,Chen2023,Grishin2023}. The gathering of abundant stars and compact objects within the inner regions of AGN disks facilitates the formation of binary systems, owing to increased encounter probability and dynamical friction in the disk \citep[e.g.,][]{Baruteau2011,Bartos2017,Yang2022}. Thus, this environment serves as a cradle for massive star explosions and mergers/collisions between stars and compact objects \citep[e.g.,][]{Cheng1999,Stone2017,Bartos2017,Yang2019,McKernan2020,Tagawa2020,Zhu2021Neutron,Zhu2021Thermonuclear,Grishin2021,Li2021,Li2022,Ren2022,LiGP2022,Li2023,Zhang2023,Luo2023}.

A key signature to identify compact binary coalescences (CBCs) embedded in the AGN accretion disks is their peculiar EM signals thanks to the surrounding dense disk medium. GRBs from the AGN disk CBCs could be choked to generate bright shock breakout emissions \citep{Zhu2021Neutron} or appear diffused \citep{Perna2021,WangYH2022}, being dependent on the burst properties, the disk structure, and the location of bursts. \cite{Yuan2022} suggested that AGN NS mergers could produce successful jets if the existence of low-density cavities and generate GeV emissions caused by the external inverse Compton processes. Bursts propagating through the disk can photo-ionize the medium to alter early-time emission \citep{Ray2023}. \cite{Lazzati2022} presented that GRBs occurring in dense environments could exhibit a single, long-emission episode because of the superposition of individual pulses, thereby transforming short-duration GRBs into long-duration ones, potentially explaining the origin of nuclear long-duration GRB\,191019A \citep{Levan2023,Lazzati2023}. Moreover, although BBH mergers are not expected to directly generate EM emission \citep[however see][]{Zhang2016}, BBH mergers within the AGN disks can be accompanied by luminous EM signals through accretion and interaction with the gaseous medium \citep{Stone2017,Bartos2017,McKernan2019,Wang2021BBH,Kimura2021,RodrguezRamirez2023,Tagawa2023}. Most intriguingly, \cite{Graham2020} reported an optical flare transient ZTF19abanrhr associated with AGN J124942.3+344929, which was the plausible EM counterpart to GW190521--a merger between two BHs with a total mass of $\sim150\,M_\odot$ \citep{Abbott2020GW190521}. 

As potential GW and EM sources, transients and CBCs embedded in the AGN disks are also expected to be ideal targets for cosmic neutrino observations. Due to the surrounding gas-rich environment, the explosions of AGN disk transients can drive powerful shocks, potentially leading to the acceleration of charged particles. High-energy neutrinos could be produced by these protons interacting with AGN disk materials and electron-radiating photons. \cite{Zhu2021HighenergyGRB} suggested that choked GRB jets inside the AGN disks are promised to allow the efficient shock acceleration of cosmic rays and produce TeV-PeV neutrinos, a process similar to that predicted for low-luminosity GRBs \citep[e.g.,][]{Murase2013,Senno2016}. Furthermore, the interactions of shock-accelerated cosmic rays with AGN disk materials after the transient ejecta shock breaks out from the AGN disk can also enable the production of high-energy neutrinos \citep{Zhu2021Highenergy}. \cite{Tagawa2023Highenergy} proposed that {protons accelerated in the internal shocks of jets launched by} solitary accreting BHs {in the open gaps of AGN disks can interact with jet photons to power high-energy} neutrino emissions, which could explain the neutrino background intensity at energies $\lesssim10^6\,{\rm GeV}$ and neutrino flux from NGC\,1068. In this {\em Letter}, we for the first time present that detectable high-energy neutrinos can be efficiently produced {in the jet head} during the propagation process of jets from AGN BBH mergers within the disk atmosphere.

\section{Formation and Propagation of Jets in Disks}

\begin{table}
 \caption{Parameters of the disk structure, accretion, and jet for $150\,M_\odot$ AGN BBH mergers locating at $10\,R_{\rm g}$, $10^3R_{\rm g}$, and $10^5R_{\rm g}$ around a $10^8M_\odot$ SMBH. }
 \label{tab:Parameter}
 \begin{tabular*}{\columnwidth}{l@{\hspace*{27pt}}l@{\hspace*{27pt}}l@{\hspace*{27pt}}l}
  \hline
  \hline
  Parameter & $10\,R_{\rm g}$ & $10^3R_{\rm g}$ & $10^5R_{\rm g}$\\[2pt]
  \hline
  \multicolumn{4}{c}{Disk and Accretion Parameters} \\[2pt]
  \hline
  $\rho_{\rm AGN}/{\rm g}\,{\rm cm}^{-3}$ & $3\times10^{-12}$ & $2\times10^{-12}$  & $1\times10^{-14}$ \\[2pt]
  $c_{s,{\rm AGN}}/{\rm cm}\,{\rm s}^{-1}$ & $3\times10^{9}$ & $1\times10^7$ & $6\times10^5$\\[2pt]
  $\Omega_{\rm AGN}/{\rm rad}\,{\rm s}^{-1}$ & $6\times10^{-5}$ & $6\times10^{-8}$ & $7\times10^{-11}$ \\[2pt]
  $H_{\rm AGN}/{\rm cm}$ & $4\times10^{13}$ & $2\times10^{14}$ & $9\times10^{15}$ \\[2pt]
  $r_{\rm BHL}/{\rm cm}$ & $3\times10^9$ & $1\times10^{14}$ & $4\times10^{16}$\\[2pt]
  $r_{\rm Hill}/{\rm cm}$ & $1\times10^{12}$ & $1\times10^{14}$ & $1\times10^{16}$\\[2pt]
  $\dot{M}_{\rm cap}/M_\odot\,{\rm yr}^{-1}$ & $9\times10^{-9}$ & $5\times10^{-2}$ & $5\times10^{-2}$\\[2pt]
  \hline
  \multicolumn{4}{c}{Jet Parameters} \\[2pt]
  \hline
  $\tilde{L}$ & $4\times10^{-4}$ & $1.1$ & $0.5$  \\[2pt]
  $L_{\rm j}/{\rm erg}\,{\rm s}^{-1}$ & $2\times10^{39}$ &  $1\times10^{46}$ & $5\times10^{46}$ \\[2pt]
  $t_{\rm j}/{\rm s}$ & $7\times10^{4}$ & $1\times10^4$ & $8\times10^{5}$ \\[2pt]
  $E_{\rm j}/{\rm erg}$ & $1\times10^{44}$ & $1\times10^{50}$ & $4\times10^{52}$ \\[2pt]
  $\beta_{\rm h}\,(\Gamma_{\rm h})$ & $0.02\,(1)$ & $0.5\,(1.2)$ & $0.4\,(1.1)$ \\[2pt]
  $\theta_{\rm j}/{\rm rad}$ & $0.006$ & $0.08$ & $0.06$ \\[2pt]
  $\tau_{\rm T}$ & $5\times10^{-5}$ & $0.4$ & $0.07$ \\[2pt]
  $\epsilon'_{p,{\rm max}}/{\rm GeV}$ & $8\times10^8$ & $9\times10^7$ & $4\times10^8$ \\[2pt]
  \hline
 \end{tabular*}
\end{table}

%\subsection{Accretion onto Remnant BHs}

Recent research has been actively exploring the accretion onto the remnants of BHs from AGN BBH mergers, as well as the properties of their associated electromagnetic EM signatures. \cite{Bartos2017} suggested that most BBH binaries can open gaps within the disk, enabling direct observations of disk outflows through super-Eddington accretions of merged BHs. However, the remnants of BBH mergers can enter the disk due to the high GW recoil kick, while the outflows have to pass through the optically thick disk, which can contribute to shock breakout signals \citep{Kimura2021}. Alternatively, BBH mergers within the AGN disks can launch powerful Blandford–Znajek jets \citep{Blandford1977} since their remnant BHs can acquire high spins and enhanced accretion rates \citep{Rossi2010,Wang2021BBH,Tagawa2023}. {{These jets are promised to directly collide with the disk materials \citep{Wang2021BBH}. Even in the presence of a gap or cavity around the BBH merger, the jets can still interact with the disk materials when the remnant BHs traverse the disk \citep{RodrguezRamirez2023} or after re-orientations of jets by the mergers \citep{Tagawa2023}.}} Figure \ref{fig:Schematic} in Appendix illustrates the physical processes of jet propagation and neutrino production for AGN BBH mergers. {For different scenarios for producing jets we described above, the accretion rate of BH can always be given by the Bondi–Holye–Lyttleton (BHL) accretion.} The capture rate of gas by the BHs is subject to the limitation of capture regions by the shear motion and the vertical height of the disk, which can be described as \citep[e.g.,][]{Tagawa2022}
\begin{equation}
\begin{split}
\label{eq:BHL}
    \dot{M}_{\rm cap} = & f_{\rm c}r_{w}r_{h}\rho_{\rm AGN}(c_{s,{\rm AGN}}^2 + v_{\rm BH}^2 + v_{\rm sh}^2)^{1/2}, \\
     \simeq & 0.14\,M_\odot\,{\rm yr}^{-1}\left(\frac{M_{\rm BH}}{150\,M_\odot}\right)^{2/3}\left(\frac{M_{\rm SMBH}}{10^8M_\odot}\right)^{-1/6} \\
     & \left(\frac{\rho_{\rm AGN}}{10^{-14}{\rm g}\,{\rm cm}^{-3}}\right)\left(\frac{R}{10^5R_{\rm g}}\right)^{1/2}\left(\frac{H_{\rm AGN}}{10^{16}{\rm cm}}\right),
\end{split}
\end{equation}
where $M_{\rm BH}$ is the remnant BH mass, $M_{\rm SMBH}$ is the SMBH mass, $f_{\rm c}=10$ is a normalization constant suggested by \cite{Tanigawa2002}, $r_w = \min(r_{\rm BHL},r_{\rm Hill})$ is the capture radius which is the minimum radius between the BHL radius $r_{\rm BHL} = GM_{\rm BH}/(c_{s,{\rm AGN}}^2 + v_{\rm BH}^2 + v_{\rm sh}^2)$ and Hill radius $r_{\rm Hill} = (M_{\rm BH}/3M_{\rm SMBH})^{1/3}R$ with the gravitational constant $G$ and the radial location $R$ of BBH binaries, $r_h = \min(r_w , H_{\rm AGN})$ is the capture height determined by $r_w$ and AGN disk scale height $H_{\rm AGN}$, $\rho_{\rm AGN}$ is the density of AGN disk surrounding the BBH binaries, $c_{s,{\rm AGN}}$ is the sound velocity of the disk at $R$, $v_{\rm BH}$ is the relative velocity between the BBH binaries and gas, $v_{\rm sh}=r_w(GM_{\rm SMBH}/R^3)^{1/2}$ is the shear velocity, and $R_{\rm g} \equiv GM_{\rm SMBH}/c^2$ is the gravitational radius of the SMBH with the speed of light $c$, respectively. The second equality of Equation (\ref{eq:BHL}) holds by assuming $v_{\rm BH}<c_{s,\rm AGN}<v_{\rm sh}$ and $r_{\rm Hill}<r_{\rm BHL}<H_{\rm AGN}$. In regard to the remnants of AGN BBH mergers, inflows rate can be enhanced with shocks arising in the outer part of the postmerger circum-BH disks caused by the GW recoil kicks impacted on the remnant \citep{Rossi2010,Tagawa2023}. Following \cite{Tagawa2023}, we introduce an enhancement factor $f_{\rm acc}$ to model the accretion rate onto the remnant BHs, i.e., $\dot{M}_{\rm BH} = f_{\rm acc}\dot{M}_{\rm cap}$. Since $f_{\rm acc}$ is an uncertain parameter, we treat it as a free parameter {{and simply study two different cases for our simulations. We assume that the accretion rate can be the BHL accretion rate and set $f_{\rm acc} = 1$. Furthermore,}} the enhancement of the accretion rate could even exceed an order of magnitude, as determined by the kick velocity and direction, {{so we assume another case with $f_{\rm acc} = 15$ similar to \cite{Tagawa2023}.}}

%\subsection{Jet Propagation in AGN Disks}

We assume that the one-side jet luminosity is given by 
\begin{equation}
\begin{split}
    L_{\rm j} =& \frac{\eta_{\rm j}\dot{M}_{\rm BH}c^2}{2} = 3\times10^{46}\,{\rm erg}\,{\rm s}^{-1}\left(\frac{\eta_{\rm j}}{0.5}\right)\left(\frac{f_{\rm acc}}{15}\right)\left(\frac{M_{\rm BH}}{150\,M_\odot}\right)^{2/3} \\
    &\left(\frac{M_{\rm SMBH}}{10^8M_\odot}\right)^{-1/6}\left(\frac{\rho_{\rm AGN}}{10^{-14}{\rm g}\,{\rm cm}^{-3}}\right)\left(\frac{R}{10^5R_{\rm g}}\right)^{1/2}\left(\frac{H_{\rm AGN}}{10^{16}{\rm cm}}\right),
\end{split}
\end{equation}
where $\eta_{\rm j}$ is the jet conversion efficiency. We set $\eta_{\rm j}\approx0.5$ for BH with a dimensionless spin of $\chi_{\rm BH}\approx0.7$ \citep[e.g.,][]{Tchekhovskoy2010,Narayan2022} for the remnant from mergers of two equal-mass, non-spinning BHs \citep[e.g.,][]{Buonanno2008}.

The jets launched from the remnant BHs would collide with AGN disk materials leading to the formation of the jet head, which is composed by a forward shock sweeping into the medium and a reverse shock entering the jet material \citep[e.g.,][]{Matzner2003,Bromberg2011}. The hot material that enters the jet head flows sideways and produces a cocoon to drive a shock inside the jet which could help collimate the jet. The jet head velocity can be expressed as \citep{Matzner2003} $\beta_{\rm h} = \beta_{\rm j}/(1 + \tilde{L}^{-1/2})$, where $\beta_{\rm j} = 1$. Thus, the jet can pass through the disk with a relativistic velocity only if $\tilde{L}\gg 1$; otherwise, the velocity would be $\beta_{\rm h}\approx\tilde{L}^{1/2}$. The critical parameter \citep[see ][for details]{Bromberg2011} that determines the jet evolution {{is defined as the dimensionless ratio between the jet’s energy density and the rest-mass energy density of AGN disk materials, i.e.,}}  $\tilde{L}\approx(L_{\rm j}/\rho_{\rm AGN}t_{\rm j}^2\theta_0^4c^5)^{2/5}$ if $\tilde{L}<\theta_0^{-4/3}$ or $\tilde{L}\approx L_{\rm j}/\rho_{\rm AGN}t_{\rm j}^2\theta_0^4c^5$ if $\tilde{L}>\theta_0^{-4/3}$, where $t_{\rm j}\approx H_{\rm AGN}/\beta_{\rm h}c$ is the jet duration before the breakout by assuming that all BBH mergers occur at the mid-plane of the disk and $\theta_0\approx0.2$ is the initial jet open angle. The jet can (cannot) be collimated by the cocoon in the former (later) case.  In our study, $\tilde{L}<\theta_0^{-4/3}$ can always be achieved, so that
\begin{equation}
\begin{split}
    \tilde{L} =& 0.9\left(\frac{\eta_{\rm j}}{0.5}\right)^{2/7}\left(\frac{f_{\rm acc}}{15}\right)^{2/7}\left(\frac{M_{\rm BH}}{150\,M_\odot}\right)^{4/21} \\
    &\left(\frac{M_{\rm SMBH}}{10^8M_\odot}\right)^{-1/21}\left(\frac{R}{10^5R_{\rm g}}\right)^{1/7}\left(\frac{H_{\rm AGN}}{10^{16}{\rm cm}}\right)^{-2/7}\left(\frac{\theta_0}{0.2}\right)^{-8/7},
\end{split}
\end{equation}
and, accordingly, the jet head velocity $\beta_{\rm h}\approx0.95(\tilde{L}/0.9)^{1/2}$, the jet duration $t_{\rm j}\approx4\times10^5{\rm s} (H_{\rm AGN}/10^{16}{\rm cm})(\tilde{L}/0.9)^{-1/2}$, and the jet open angle $\theta_{\rm j} \approx\tilde{L}^{1/4}\theta_0^2=0.04(\tilde{L}/0.9)^{1/4}(\theta_0/0.2)^2$. The jet energy $E_{\rm j}=L_{\rm j}t_{\rm j}$ before the breakout can be expressed as
\begin{equation}
\begin{split}
\label{eq:JetEnergy}
    E_{\rm j} &= 10^{52}{\rm erg}\left(\frac{\eta_{\rm j}}{0.5}\right)^{6/7}\left(\frac{f_{\rm acc}}{15}\right)^{6/7}\left(\frac{M_{\rm BH}}{150\,M_\odot}\right)^{4/7}\left(\frac{M_{\rm SMBH}}{10^8M_\odot}\right)^{-1/7} \\
    &\left(\frac{\rho_{\rm AGN}}{10^{-14}{\rm g}\,{\rm cm}^{-3}}\right)\left(\frac{R}{10^5R_{\rm g}}\right)^{3/7}\left(\frac{H_{\rm AGN}}{10^{16}{\rm cm}}\right)^{15/7}\left(\frac{\theta_0}{0.2}\right)^{4/7}.
\end{split}
\end{equation}
Thus, the jet energy is mainly determined by the remnant BH mass, the SMBH mass, and the properties of disk at the position of the BBH merger. The parameter with the most influence on the jet energy is the disk scale height around the BBH merger. Thus, BBH mergers occurring in the outer regions of accretion disks around massive SMBHs, where there is typically a larger disk height, are more likely to generate more powerful jets, potentially exceeding classical GRBs energies \citep[e.g.,][]{Zhang2018}. As we will shown in Section \ref{sec:NeutrinoProduction}, the neutrino fluence highly depends on the jet energy. One can thus expect that a more powerful neutrino burst for a BBH merger occurring at more outer part of the disk around massive SMBHs.

Based on the disk structure modelled by \cite{Thompson2005}, we study the properties of neutrino bursts from AGN BBH mergers for specific models in Section \ref{sec:NeutrinoProduction}. We take GW190521-like BBH merger with a total mass of $150\,M_\odot$ around a $10^8M_\odot$ SMBH at three different radial distances $10\,R_{\rm g}$, $10^3R_{\rm g}$, and $10^5R_{\rm g}$ as examples. By adopting $\eta_{\rm j}=0.5$ and $f_{\rm acc} = 15$, the parameters of the disk structure, accretion, and jet for these three models are summarized in Table \ref{tab:Parameter}. It is expected that these jets from AGN BBH mergers can be always highly collimated with a non-relativistic or mildly-relativistic motion inside the accretion disks. Furthermore, the jet energy could have a very wide distribution depending on the location of the BBH mergers. The dependence of the accretion parameter and remnant BH mass on the detection of neutrino bursts from AGN BBH mergers will be discussed in Section \ref{sec:Properties}.

\section{Cosmic Rays and Neutrino Production} \label{sec:NeutrinoProduction}

\subsection{Maximum Proton Energy}

\begin{figure*}
     \centering
     \includegraphics[totalheight=2.03in , trim = 52 25 92 59, clip]{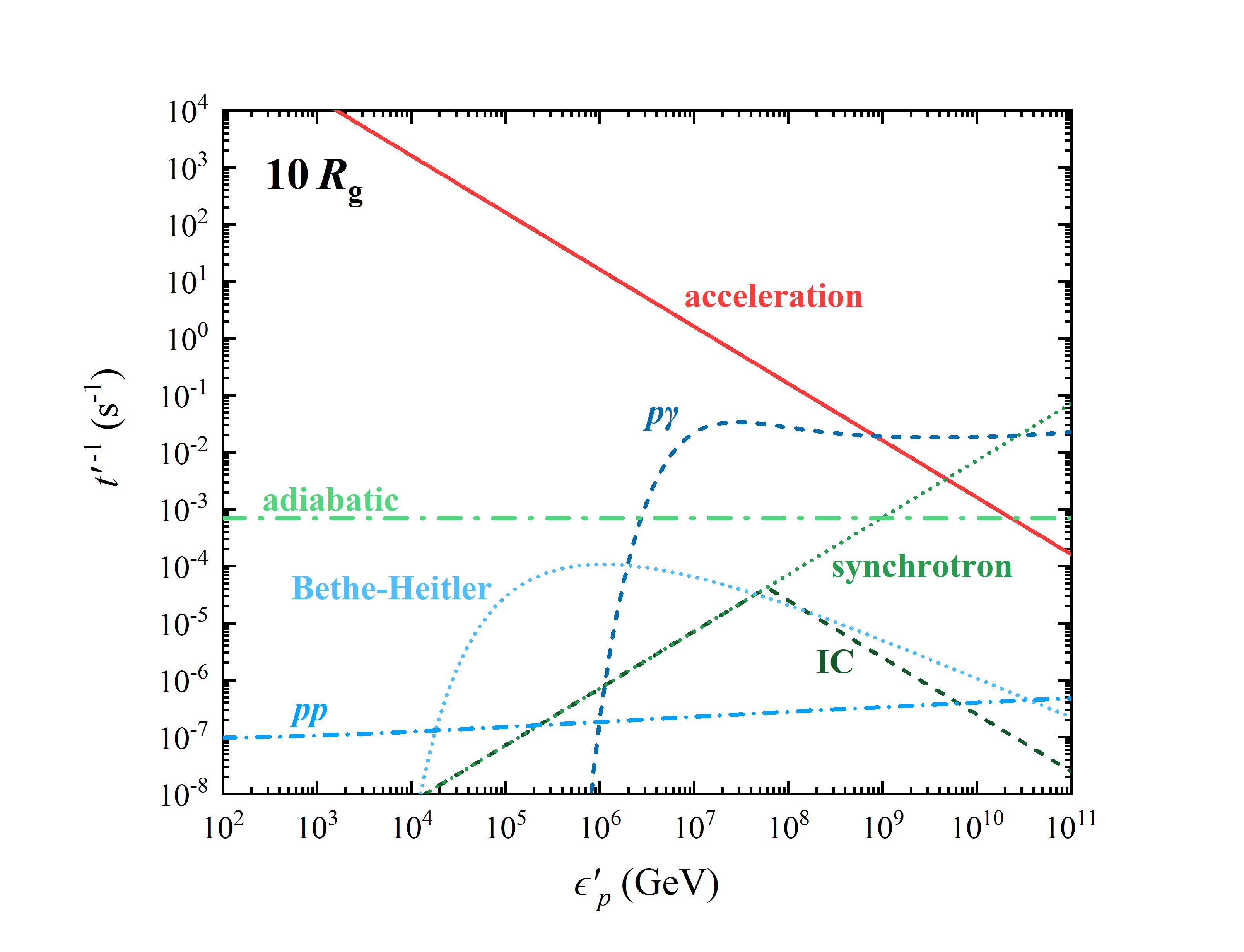}
     \includegraphics[totalheight=2.03in , trim = 126 25 92 59, clip]{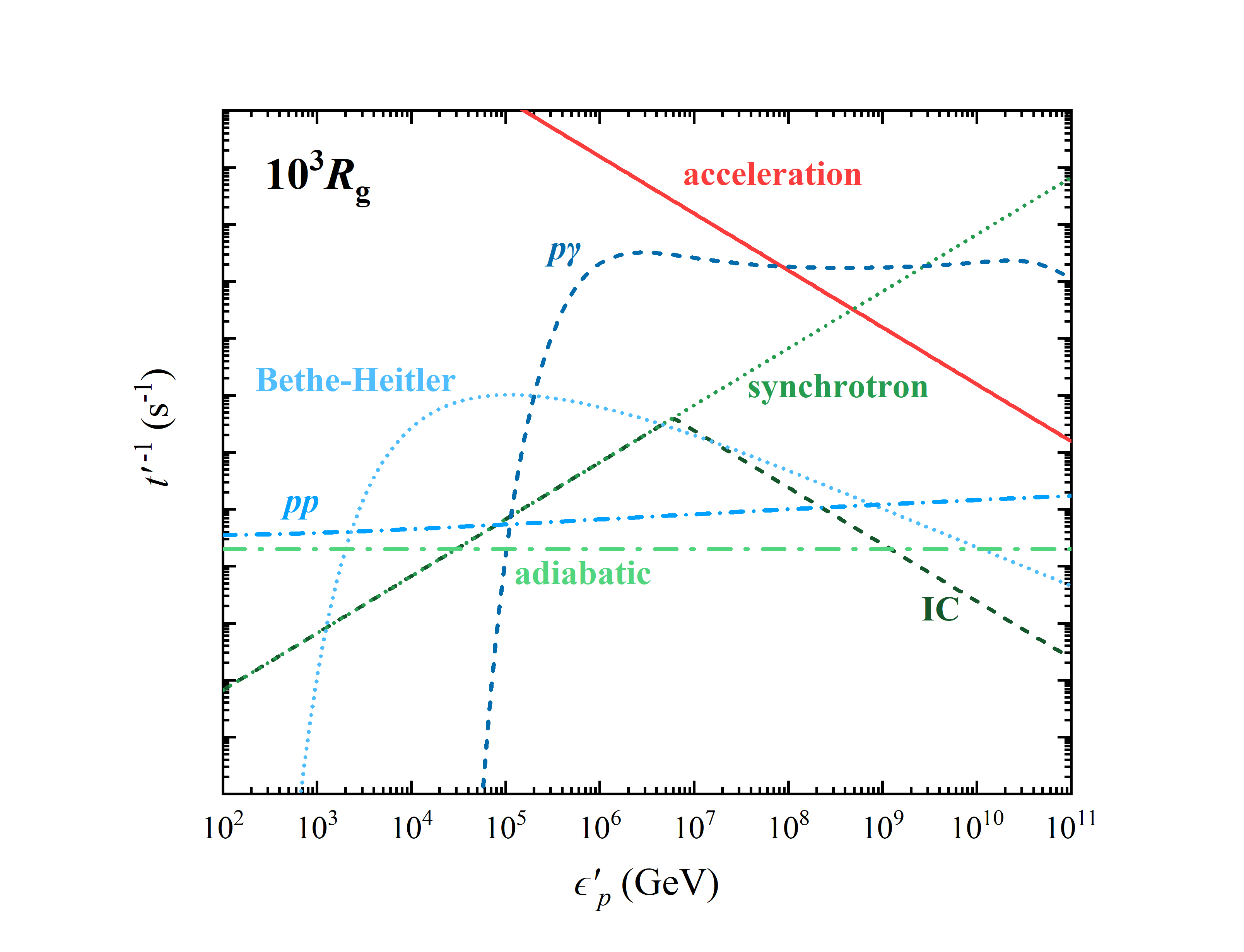}
     \includegraphics[totalheight=2.03in , trim = 126 25 92 59, clip]{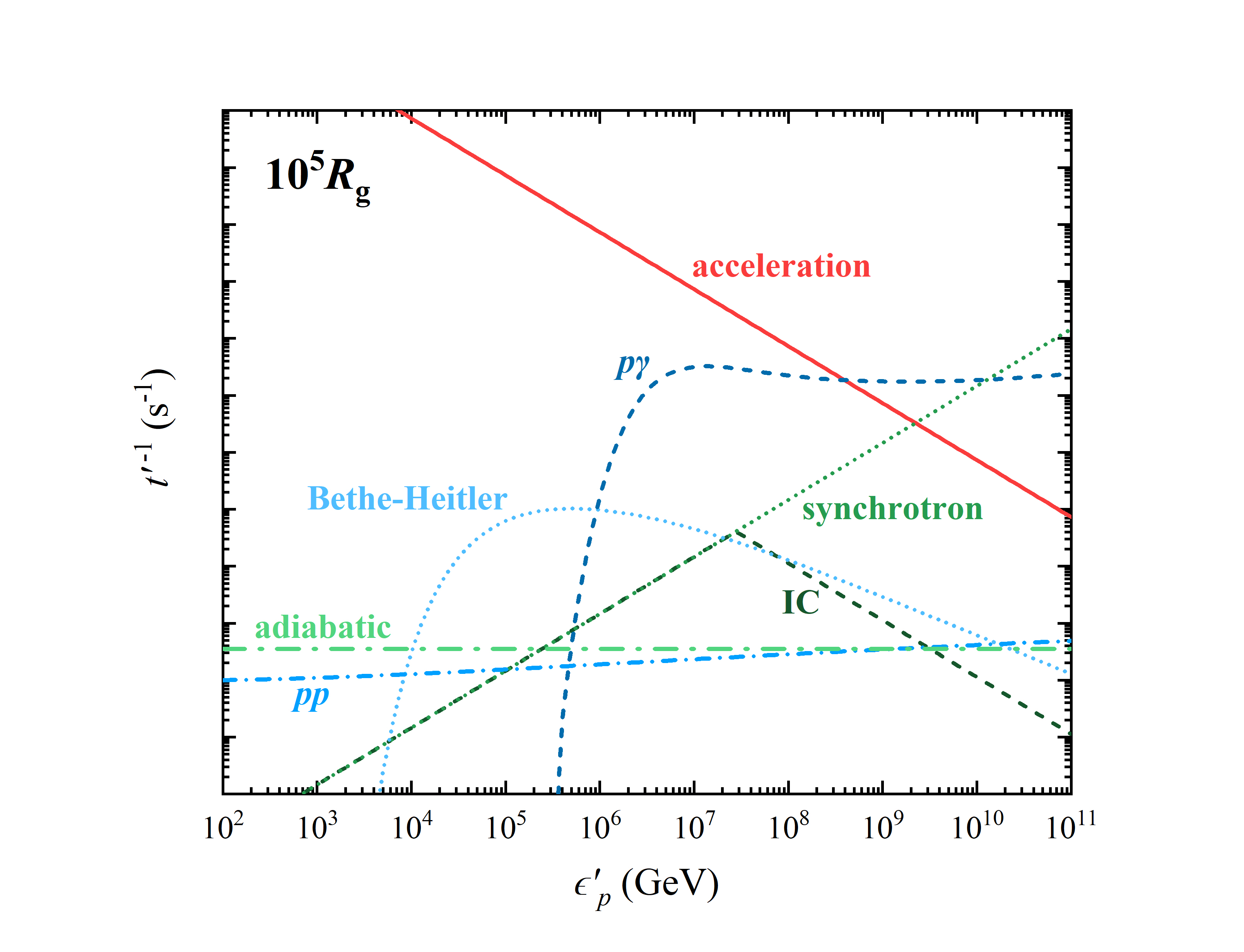}
     \caption{Inverse of proton acceleration and cooling timescales as a function of proton energy in the jet head frame for $150\,M_\odot$ remnant BHs occurring at the radial distances of $10\,R_{\rm g}$ (left panel), $10^3R_{\rm g}$ (middle panel) and $10^5R_{\rm g}$ (right panel) around a $10^8M_\odot$ SMBH. Acceleration (red line), photomeson production ($p\gamma$, blue dashed), Bethe–Heitler pair production (blue dotted), hadronuclear scattering ($pp$, blue dashed–dotted), inverse-Compton (IC, green dashed), synchrotron radiation (green dotted), and adiabatic cooling (green dashed–dotted) processes are considered.}
     \label{fig:timescale}
\end{figure*} 

We assume that the Fermi acceleration operates and the accelerated protons have a power-law number density distribution with a highest-energy exponential cutoff in energy: $dn'_p/d\epsilon'_p{\propto}{\epsilon'_p}^{-2}\exp({-\epsilon'_p/\epsilon'_{p,{\rm max}}}),$ where the maximum proton energy $\epsilon'_{p,{\rm max}}$ is determined by the proton acceleration timescale and cooling timescales. We define three reference frames: $Q$ for the AGN rest frame, $Q'$ for the jet head comoving frame, and $Q''$ for the jet comoving frame.

The internal energy and density evolution of the forward shock and reverse shock in the jet head can be described by the shock jump conditions \citep{Blandford1976,Sari1995,Zhang2018}, i.e.,
\begin{equation}
\begin{split}
\label{eq:ShockJumpCondition}
    e'_{\rm f}/n'_{\rm f}m_pc^2 &= \Gamma_{\rm h} - 1,~n'_{\rm f}/n_{\rm a} = (\hat{\gamma}_{\rm 1}\Gamma_{\rm h} + 1)/(\hat{\gamma}_{\rm 1} - 1)=4\Gamma_{\rm h},\\
    e'_{\rm r}/n'_{\rm r}m_pc^2 &= \bar{\Gamma}_{\rm h} - 1,~n'_{\rm r}/n''_{\rm j} = (\hat{\gamma}_{\rm 2}\bar{\Gamma}_{\rm h} + 1)/(\hat{\gamma}_{\rm 2} - 1)=4\bar{\Gamma}_{\rm h},
\end{split}
\end{equation}
where the subscripts ``a'', ``f'', ``r'', and ``j'' represent regions of the unshocked AGN disk medium, the jet head's forward shock, the jet head's reverse shock, and the unshocked jet, $\hat{\gamma}_1=(4\Gamma_{\rm h}+1)/3\Gamma_{\rm h}$, $\hat{\gamma}_2=(4\bar{\Gamma}_{\rm h}+1)/3\bar{\Gamma}_{\rm h}$, $\Gamma_{\rm h}$ is the Lorentz factor of the jet head, $\Gamma_{\rm j}\approx1/\theta_0$ is the Lorentz factor of the unshocked jet, and $\bar{\Gamma}_{\rm h} = \Gamma_{\rm j}\Gamma_{\rm h}(1-\beta_{\rm j}\beta_{\rm h})$ is the Lorentz factor of the unshocked jet measured in the jet head frame. Following Equation (\ref{eq:ShockJumpCondition}), the proton energy density of the jet head can be expressed as $n'_p{\approx}n'_{\rm r}=4\bar{\Gamma}_{\rm h}n''_{\rm j}$, where the jet density is $n''_{\rm j}=L_{\rm j}/\pi\theta_{\rm j}^2H_{\rm AGN}^2m_{p}\Gamma_{\rm j}^2c^3$ with the proton mass $m_p$. The photon temperature in the jet head is $k_{\rm B}T'_{\rm r} = (15\hbar^3c^3 \varepsilon_ee'_{\rm r}/\pi^2)^{1/4}$, where $k_{\rm B}$ is the Boltzmann constant, $\hbar$ is the reduced Planck constant, and the electron energy fraction is set to be $\varepsilon_{e}\approx0.1$ frequently used in the literature. One can thus obtain the average thermal photon energy $\epsilon'_\gamma = 2.7k_{\rm B}T'_{\rm r}$ and the average thermal photon density $n'_\gamma = \varepsilon_ee'_{\rm r}/\epsilon'_\gamma$.

{{As was pointed out in \cite{Murase2013}, the velocity gain in each shock of protons will not be enough to achieve Fermi acceleration when the jet density is excessively high, thereby preventing the acceleration of particles to high energy. We apply the constraint on the Thomson optical depth of the jet $\tau''_{\rm T}{\approx}n''_{\rm j}\sigma_{\rm T}H_{\rm AGN}/\Gamma_{\rm j}\lesssim1$ to ensure whether protons are accelerated successfully, where $\sigma_{\rm T}$ is the Thomson cross section. As listed in Table \ref{tab:Parameter}, all three example models meet this constraint, indicating that Fermi acceleration is efficient. }}

The acceleration timescale in the jet head frame is given by $t'_{p,\rm acc}=\epsilon'_p/eB'c$, under the assumption of perfectly efficient acceleration \citep[e.g.,][]{Rachen1998}, where $e$ is the electron charge and the magnetic field strength is $B'\approx\sqrt{8\pi\varepsilon_{B}e'_{\rm r}}$ with the magnetic field energy fraction adopted as the common value $\varepsilon_{B}\approx0.1$. 

A high-energy proton can lose its energy via radiative, hadronic, and adiabatic cooling processes. The radiative cooling mechanisms include synchrotron radiation with cooling timescale of $t'_{p,\rm syn} = {6\pi m_p^4c^3}/{\sigma_{\rm T}m_e^2B'^2\epsilon_p'},$ and inverse-Compton scattering with cooling timescale of $t'_{p,\rm IC} ={3m_p^4c^3}/{4\sigma_{\rm T}m_e^2{n}'_\gamma{\epsilon}'_\gamma\epsilon_p'}$ if ${\epsilon}'_\gamma\epsilon_p'<m_p^2c^4$ and $t'_{p,\rm IC} =3\epsilon_\gamma'\epsilon_p'/{4\sigma_{\rm T}m_e^2c^5{n}'_\gamma}$ if ${\epsilon}'_\gamma\epsilon_p'>m_p^2c^4$, where $m_{e}$ is the electron mass. The hadronic cooling mechanisms mainly contain the inelastic hadronuclear scattering ($pp$), the Bethe-Heitler pair production ($p\gamma{\rightarrow}pe^+e^-$), and the photomeson production ($p\gamma$). While  charged pions/kaons produced by $pp$ and $p\gamma$ scatterings can further decay to give rise to astrophysical neutrinos, the Bethe-Heitler production would suppress the formation of high-energy neutrinos. The cooling timescale of $pp$ scattering is $t'_{p,pp}= {1}/{\kappa_{pp}\sigma_{pp}n'_pc},$ where the inelasticity is $\kappa_{pp}\simeq0.5$ and the cross section is given by \cite{Kelner2006}, i.e. $\sigma_{pp}=(34.3+1.88L+0.25L^2)(1-(\epsilon'_p/\epsilon_{\rm th})^4)^2{\rm mb}$ with $L=\ln(\epsilon'_p/1\,{\rm TeV})$, the threshold energy of production of pions $\epsilon_{\rm th}=m_p+2m_\pi+m_\pi^2/m_p$ and the pion mass $m_\pi$. We use the formula given by \cite{Stecker1968} and \cite{Murase2007} to calculate the energy loss rate of $p\gamma$ production
\begin{equation}
\label{eq:pgamma}
    t'^{-1}_{p,p\gamma} = \frac{c}{2\gamma_p}\int^\infty_{\bar{\epsilon}_{\rm th}}d\bar{\epsilon}\sigma_{p\gamma}(\bar{\epsilon})\kappa_{p\gamma}(\bar{\epsilon})\bar{\epsilon}\int^\infty_{\bar{\epsilon}/2\gamma_p}d\epsilon\epsilon^{-2}\frac{dn}{d\epsilon},
\end{equation}
where $\gamma_p=\epsilon'_p/m_pc^2$, $\bar{\epsilon}$ is the photon energy in the rest frame of the proton, $\bar{\epsilon}_{\rm th}\approx145\,{\rm MeV}$ is the threshold energy, and $dn/d\epsilon$ is the photon number density in the energy range of $\epsilon$ to $\epsilon+d\epsilon$. The inelasticity $\kappa_{p\gamma}$ and the cross section $\sigma_{p\gamma}$ are taken from \cite{Stecker1968} and \cite{Patrignani2016}, respectively. Furthermore, we obtain the energy loss rate of the Bethe-Heitler production $t'^{-1}_{\rm BH}$ by substituting $\bar{\epsilon}_{\rm th}$, $\kappa_{\rm BH}$, and $\sigma_{\rm BH}$ collected from \cite{Chodorowski1992} into $\bar{\epsilon}_{\rm th}$, $\kappa_{p\gamma}$, and $\sigma_{p\gamma}$ in Equation (\ref{eq:pgamma}). Finally, the protons are subject to the adiabatic cooling which gives a cooling timescale of $t'_{p,{\rm adi}}\approx{H_{\rm AGN}}/{c\Gamma_{\rm h}}.$

We display the acceleration and cooling timescales of our example models in Figure \ref{fig:timescale}. The $pp$ scattering can be partially suppressed by the adiabatic cooling for low-energy protons if BBH mergers occur in the outer part of AGN accretion disks, while the inner-side AGN BBH mergers would lead to a significant energy loss rate of adiabatic caused by the small disk height to dominate the cooling mechanism of low-energy protons. The Bethe-Heitler process could suppress neutrino production if the energy of protons falls within $1\,{\rm TeV}\lesssim\epsilon'_p\lesssim10^3\,{\rm TeV}$. At higher energies, $p\gamma$ scattering can always be the dominant cooling process for protons, which also determines the maximum proton energy to $\sim10^2-10^3\,{\rm PeV}$.

\subsection{Neutrino Production}

\begin{figure}
     \centering
     \includegraphics[width=1\linewidth , trim = 58 35 92 59, clip]{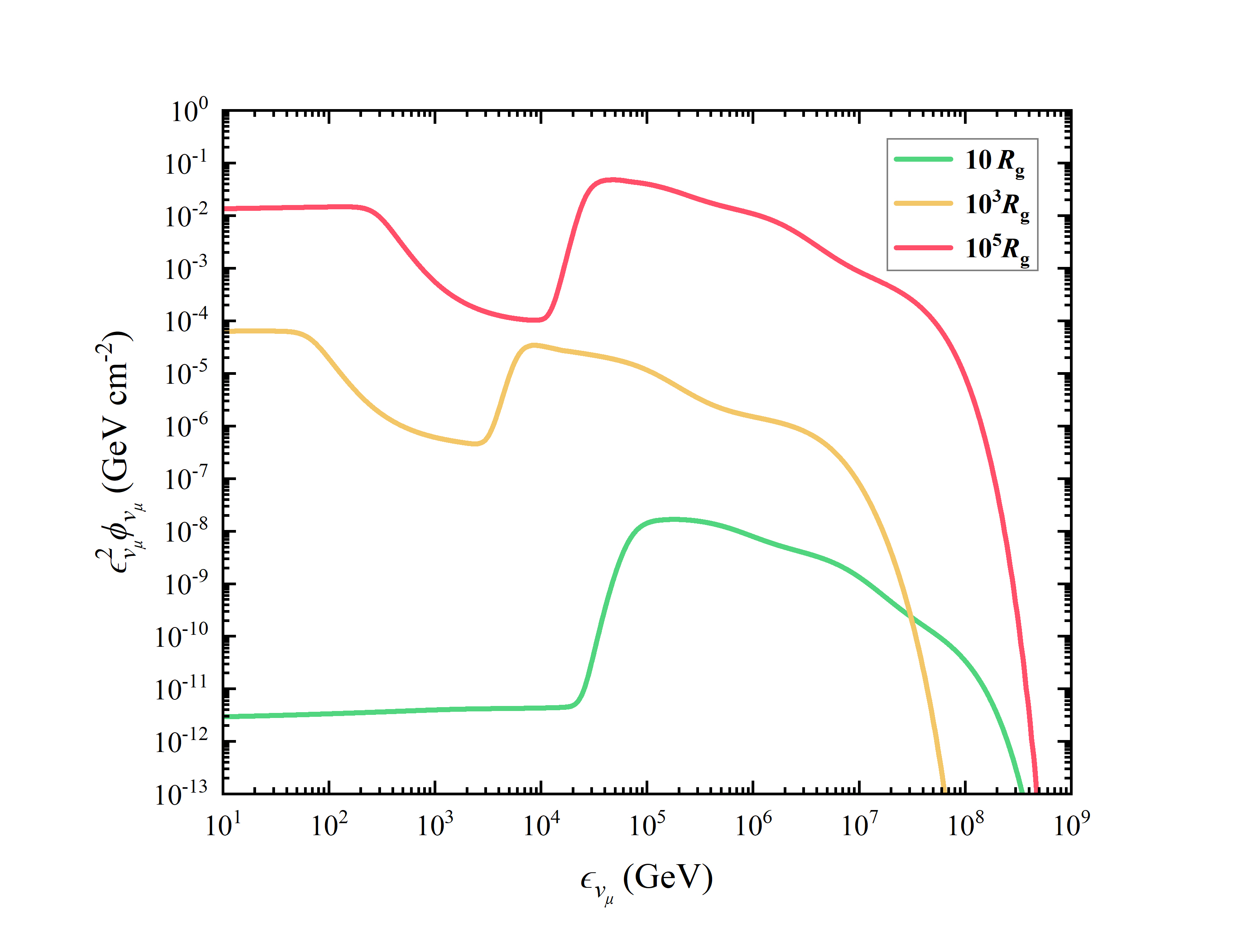}
     \caption{Expected muon neutrino fluences as a function of neutrino energy for on-axis jets from AGN BBH mergers with a total mass of $150\,M_\odot$ at $D_{\rm L} = 5300\,{\rm Mpc}$. Purple, yellow, and green lines are for BBH mergers at the radial distances of $10\,R_{\rm g}$, $10^3R_{\rm g}$, and $10^5R_{\rm g}$ around a $10^8M_\odot$ SMBH, respectively.  }
     \label{fig:fluence}
\end{figure} 

Charged pions and kaons produced in $pp$ and $p\gamma$ interactions decay into muons and muon neutrinos. Charged pions and kaons can also experience hadronic scattering whose cooling timescale is $t'_{\{\pi,K\},\pi p}=1/\kappa_{\pi p}\sigma_{\pi p}n'_pc$, where the inelasity $\kappa_{\pi p}\approx0.8$ and the pion-proton scattering cross section $\sigma_{\pi p}\approx5\times10^{-26}{\rm cm}^2$ \citep{Olive2014}. The intermediate muons then further decay to muon neutrinos, electron neutrinos, and electrons. Similar to protons, pions, kaons, and muons are also subject to radiative and adiabatic cooling processes. The energy fractions from a proton to intermediate particles are calculated based on \cite{Tamborra2016}. We get the final neutrino spectrum by comparing these cooling timescales with the decay timescale of intermediate particles, i.e., $t_{i,{\rm dec}}=\gamma_i\tau_i$, where $\gamma_i=\epsilon'_i/m_ic^2$ is the Lorentz factor, $\tau_i$ is the rest frame lifetime, and $i=\pi,\ K,\ \mu_\pi$, and $\mu_K$ are the parent particles of the neutrinos.

Because high-energy neutrinos can only be generated via $pp$ and $p\gamma$ interactions and can be suppressed by other proton cooling process, the suppression factor taking into account various proton cooling processes \citep[e.g.,][]{Murase2008,Wang2009,Xiao2014,Xiao2016} is written as $\zeta_{p,\rm sup}({\epsilon_{\nu,i}})=(t'^{-1}_{pp} + t'^{-1}_{p\gamma})/t'^{-1}_{p,\rm cool}$, where the inverse of the total cooling for protons is $t'^{-1}_{p,{\rm cool}}=t'^{-1}_{p,pp} + t'^{-1}_{p,p\gamma} + t'^{-1}_{p,{\rm BH}} + t'^{-1}_{p,{\rm syn}} + t'^{-1}_{p,{\rm IC}} + t'^{-1}_{p,{\rm adi}}$. Similarly, the suppression factor for the meson cooling can be expressed as $\zeta_{i,{\rm sup}}({\epsilon_{\nu,i}})=t'^{-1}_{i,\rm dec}/t'^{-1}_{i,\rm cool}$. The inverse of cooling timescales for pions and kaons is $t'^{-1}_{i,\rm cool} = t'^{-1}_{i,\rm dec} + t'^{-1}_{i,\pi p} + t'^{-1}_{i,\rm syn} + t'^{-1}_{i,\rm IC} + t'^{-1}_{i,\rm adi}$, while the hadronic cooling process need not be considered for the muon cooling. One can thus obtain the neutrino fluence in each neutrino production channel for a single event at a luminosity distance of $D_{\rm L}$, i.e.,
\begin{equation}
    \epsilon^2_{\nu_i}\phi_{\nu_i}\approx\frac{N_iE_{\rm iso}\zeta_{p,{\rm sup}}({\epsilon_{\nu_i}})\zeta_{i,{\rm sup}}({\epsilon_{\nu_i}})\exp(\epsilon'_p/\epsilon'_{p,\rm max})}{4\pi D_{\rm L}^2\ln(\epsilon'_{p,{\max}}/\epsilon'_{p,{\min}})},
\end{equation}
where $N_\pi=N_{\mu_\pi}=0.12$, $N_K=0.009$, $N_{\mu_K}=0.003$, the neutrino energy is $\epsilon_{\nu_i}=\Gamma_{\rm h}a_i\epsilon'_p$ with $a_\pi = a_{\mu_\pi}=0.05$, $a_K=0.10$ and $a_{\mu_K}=0.10$ \citep[e.g.,][]{Tamborra2016} for $p\gamma$ scattering, $E_{\rm iso}\approx4L_{\rm j}t_{\rm j}/\theta_{\rm j}^2$ is the isotropic equivalent energy, and $\ln(\epsilon'_{p,{\max}}/\epsilon'_{p,{\min}})$ is the normalized factor with $\epsilon'_{p,{\rm min}}\sim\Gamma_{\rm h}m_pc^2$.

Muon neutrinos and electron neutrinos produced in $pp$ and $p\gamma$ interactions change their flavor during the propagation to the Earth. In Figure \ref{fig:fluence}, we show the expected muon neutrino fluence of our example AGN BBH mergers after considering the redshift effect and neutrino oscillation \citep[e.g.,][]{Harrison2002}. We set $D_{\rm L} = 5300\,{\rm Mpc}$, corresponding to a redshift of $z\sim0.82$, as a reference value, which is the measured distance of GW190521 \citep{Abbott2020GW190521}. The fluence is mainly determined by the energy of the jets. The dip around $\sim10\,{\rm TeV}$ is resulted from the suppression the suppression of the neutrino fluence by the Bethe-Heitler production. One can see that low-energy neutrinos is dominated by the $pp$ scattering, while $p\gamma$ scattering contributes to the neutrinos above $\sim10\,{\rm TeV}$ that we are interested in.

\section{Parameter Dependence of Neutrino Burst Detection} \label{sec:Properties}

\begin{figure*}
     \centering
     \includegraphics[totalheight=2.21in , trim = 44 92 217 29, clip]{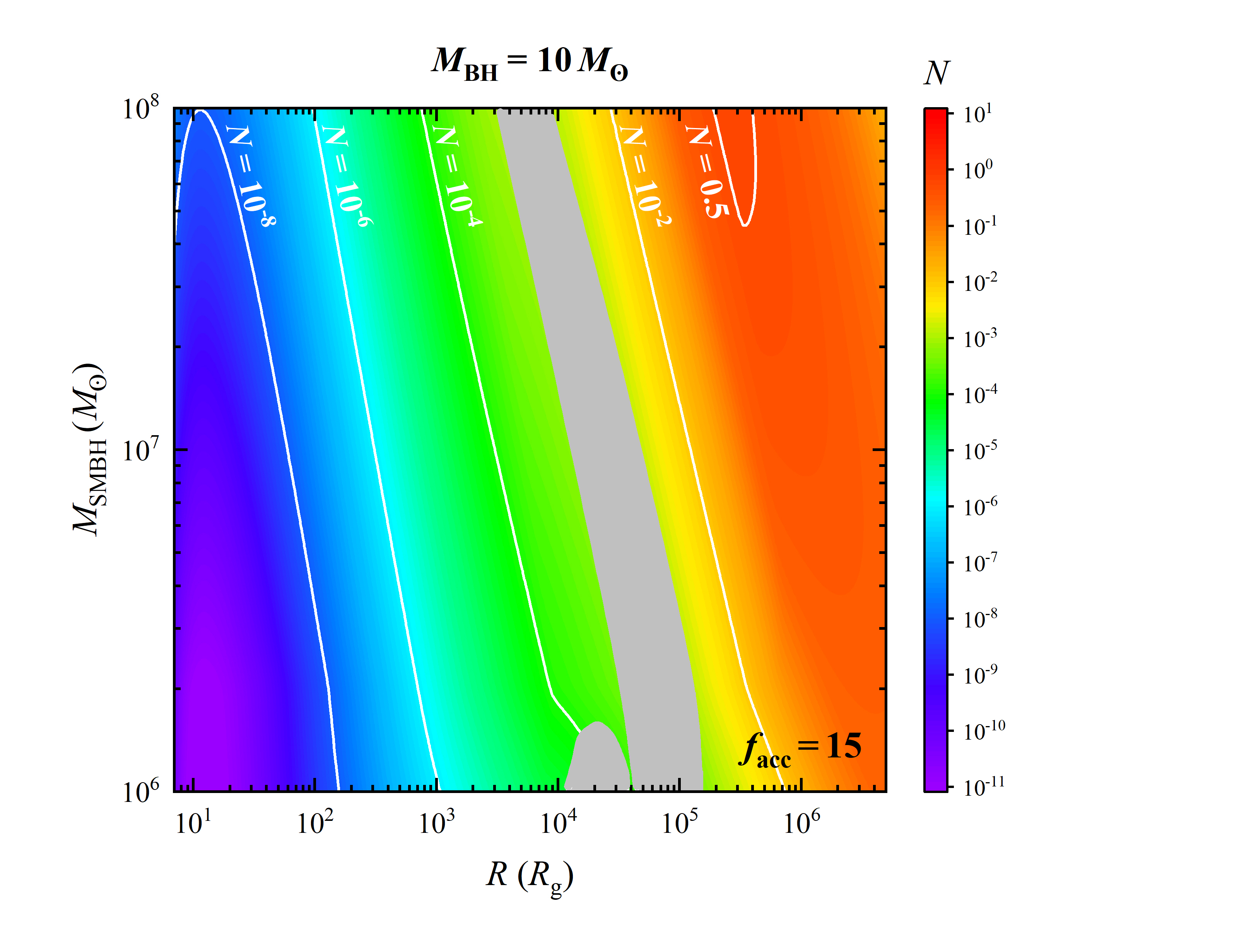}
     \includegraphics[totalheight=2.21in , trim = 105 92 217 29, clip]{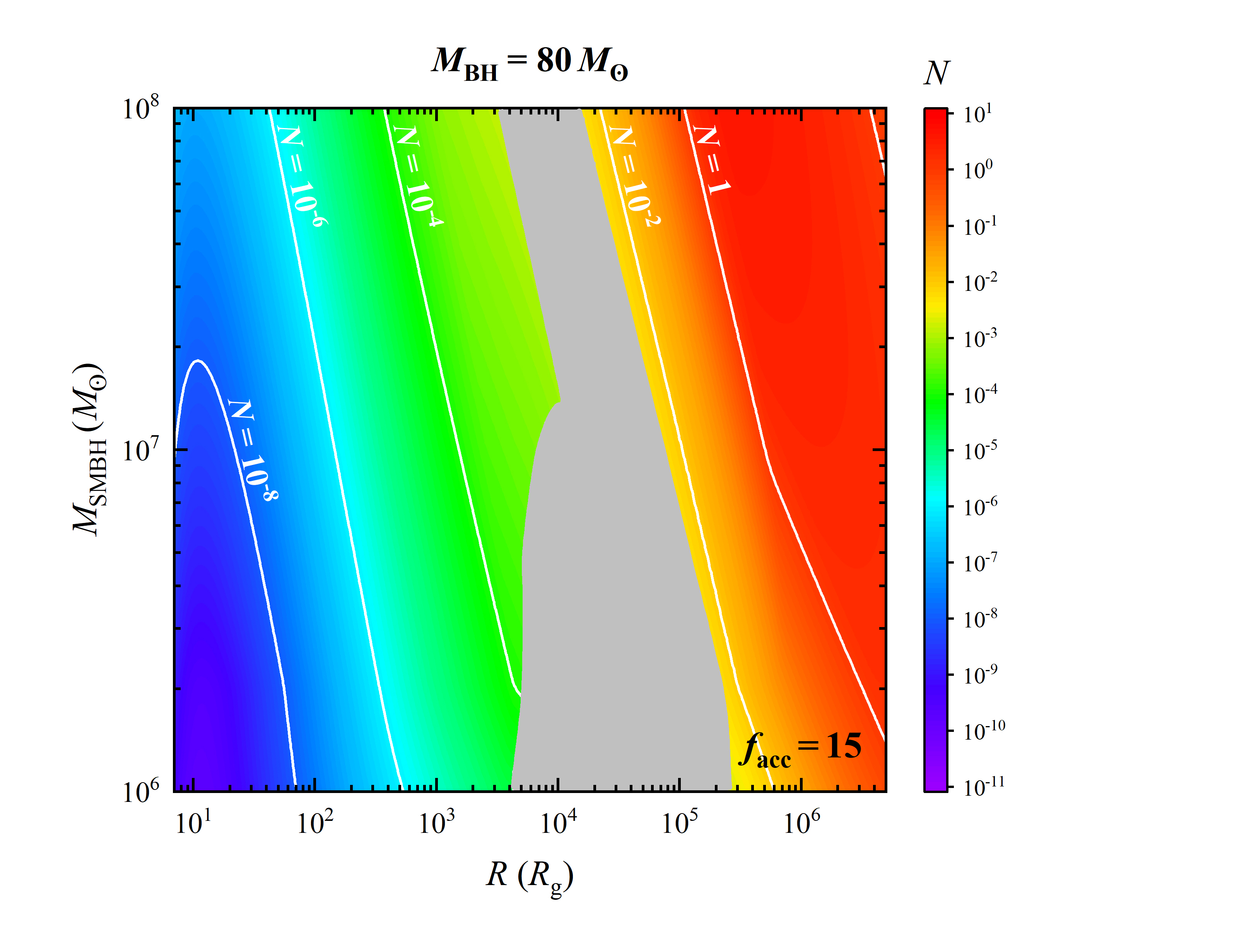}
     \includegraphics[totalheight=2.21in , trim = 105 92 148 29, clip]{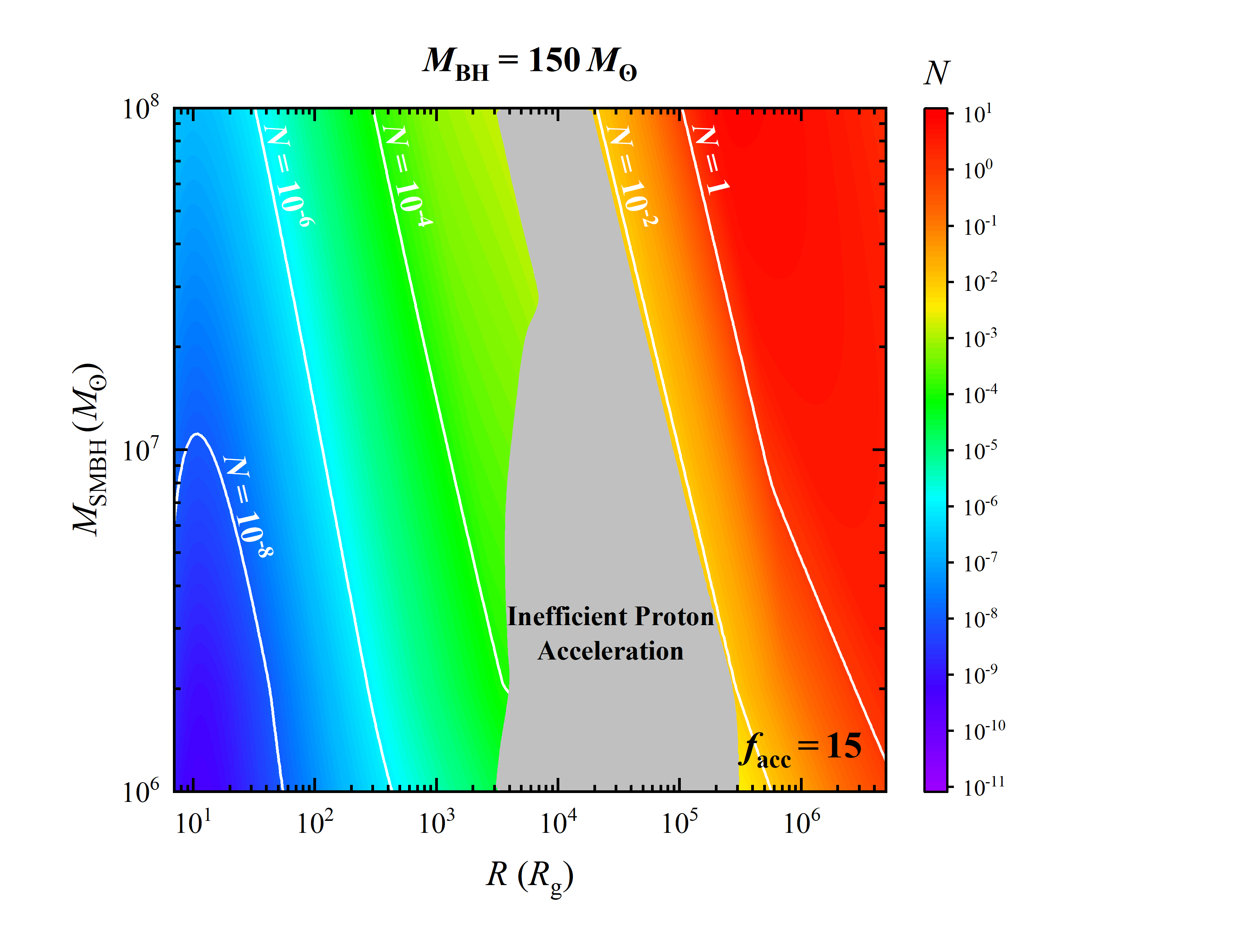}
     \includegraphics[totalheight=2.3575in , trim = 44 32 217 58, clip]{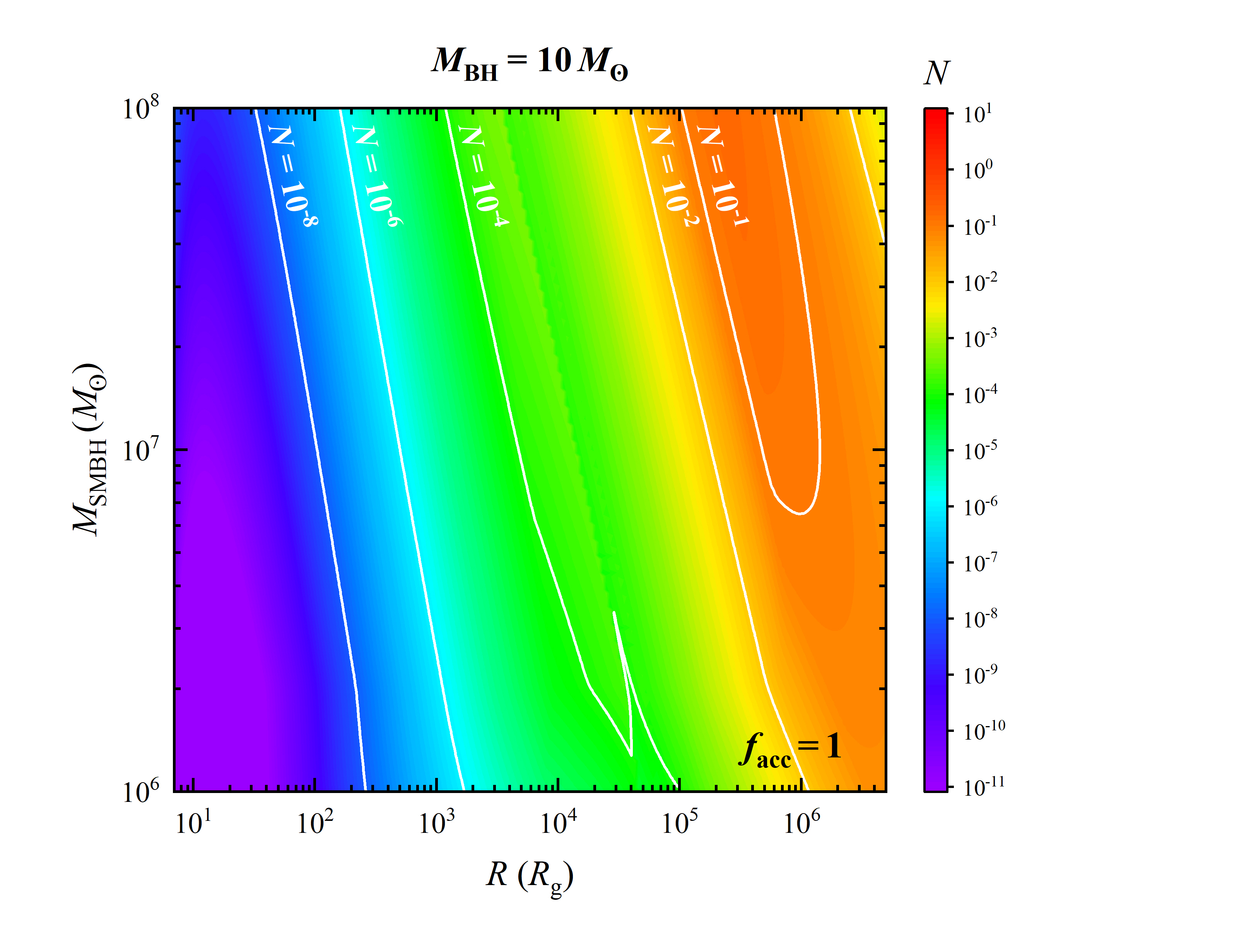}
     \includegraphics[totalheight=2.3575in , trim = 105 32 217 58, clip]{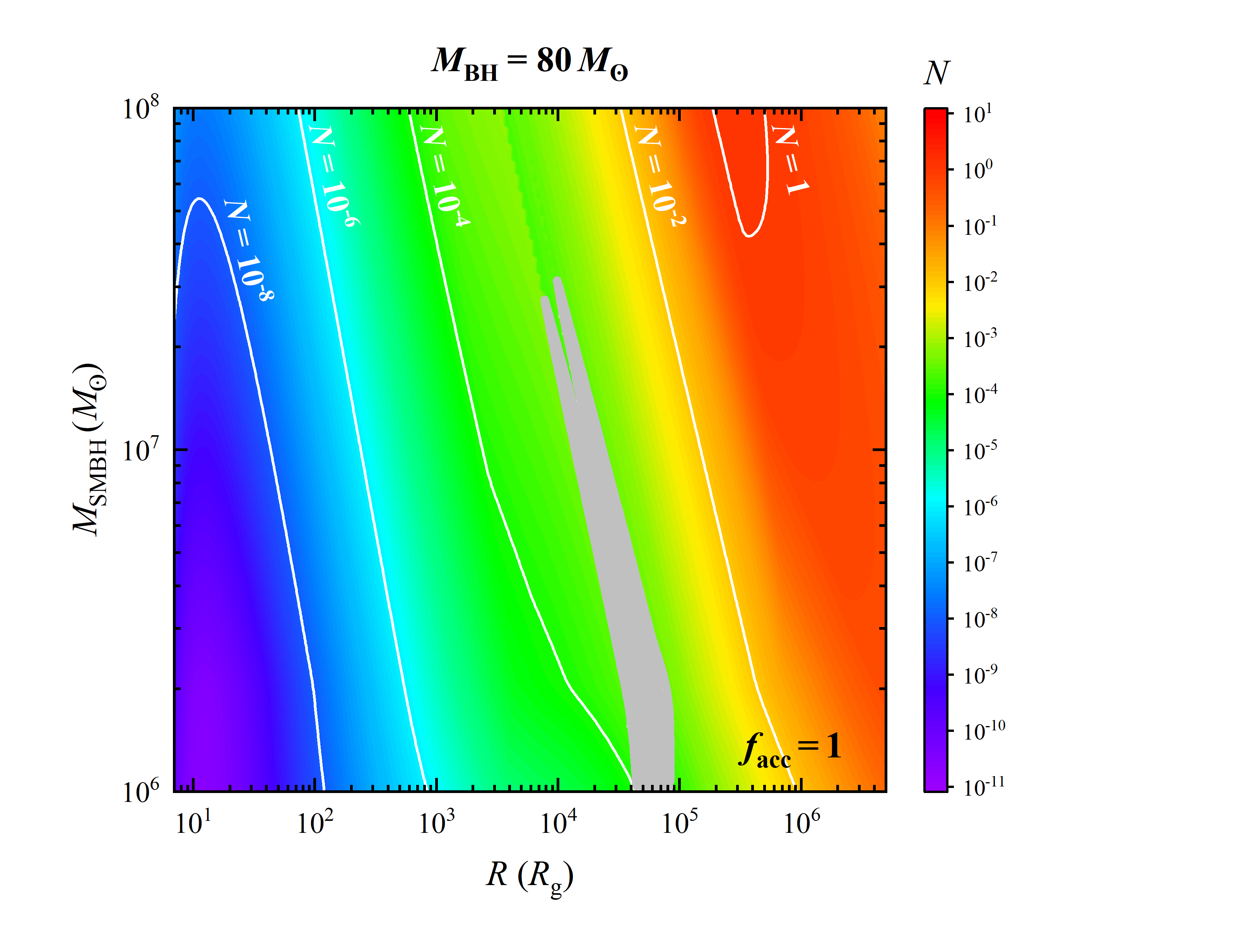}
     \includegraphics[totalheight=2.3575in , trim = 105 32 148 58, clip]{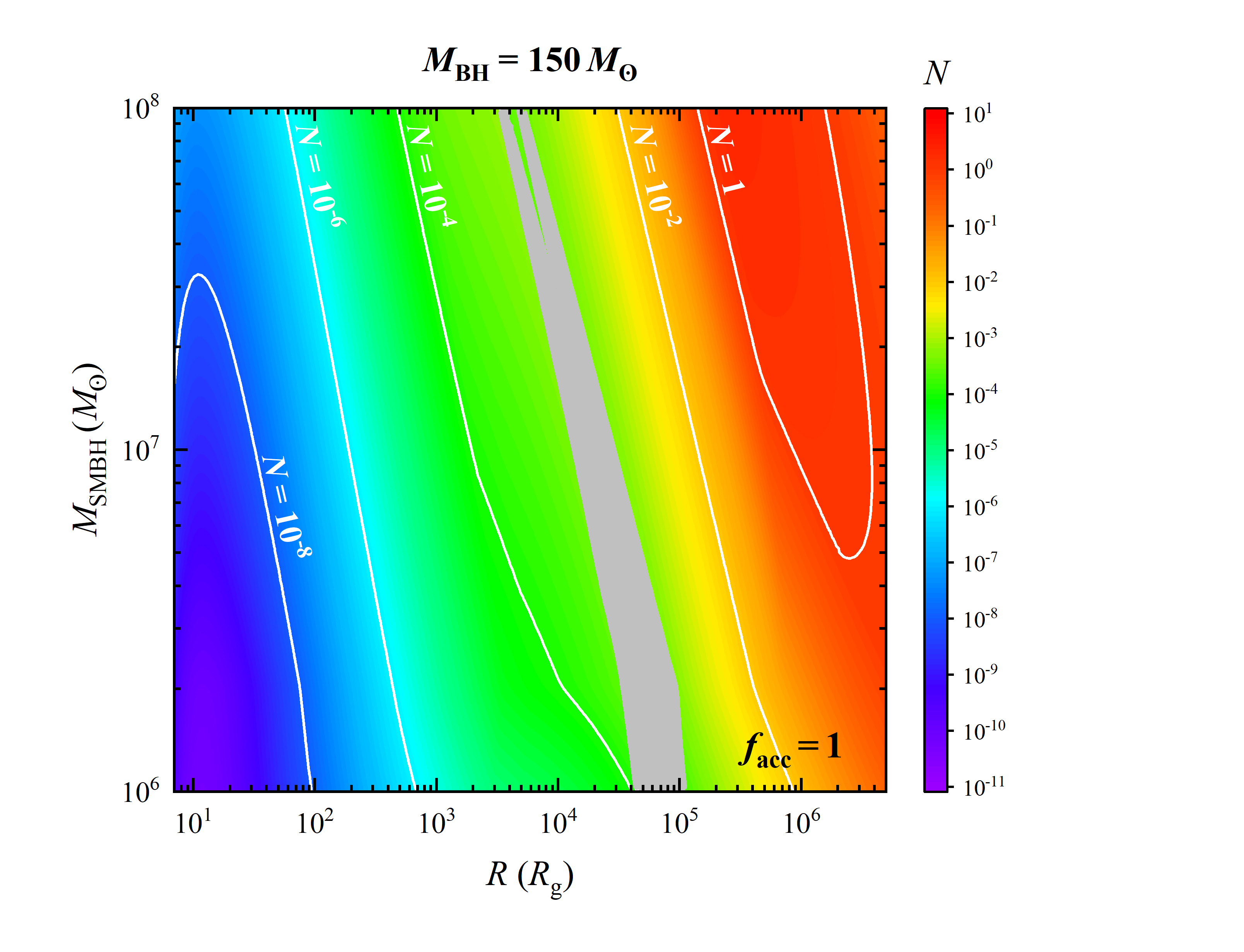}
     \caption{Parameter space in the $R-M_{\rm SMBH}$ plane with color indicating the expected number of astrophysical muon neutrinos from on-axis jets of AGN BBH mergers detectable by IceCube. Three different total masses of AGN BBH mergers, including $M_{\rm BH}=10\,M_\odot$ (left panels), $80\,M_\odot$ (middle panels), and $150\,M_\odot$ (right panels), and two different enhancement factors, i.e., $f_{\rm acc}=15$ (top panels) and $f_{\rm acc}=1$ (bottom panels) are considered. We mark several values of the expected number of muon neutrinos as solid lines in each panel. Grey shadows represent the forbidden region that for proton acceleration. }
     \label{fig:Number}
\end{figure*} 

We calculate the expected number of astrophysical muon neutrinos from on-axis jets of AGN BBH mergers detectable by IceCube following $N(\epsilon_{\nu_\mu}>1\,{\rm TeV}) = \int^{\epsilon_{{\nu_\mu},\rm max}}_{1\,{\rm TeV}}d\epsilon_{\nu_\mu}\phi_{\nu_\mu}(\epsilon_{\nu_\mu})A_{\rm eff}(\epsilon_{\nu_\mu}),$ where $A_{\rm eff}$ is the effective area of the detector. Hereafter, we only consider up-going neutrino events. The effective area of IceCube is obtained from \cite{Aartsen2017}.

As illustrated in Figure \ref{fig:Number}, we show the number of detected muon neutrinos from a single event located at $5300\,{\rm Mpc}$. Our analysis includes three specific remnant BH masses: $10\,M_\odot$, $80\,M_\odot$, and $150\,M_\odot$, along with two different enhancement factors, i.e., $f_{\rm acc}=15$ and $f_{\rm acc}=1$. Since the neutrino fluence is determined by the total jet energy, which is primarily influenced by the disk scale height as described in Equation (\ref{eq:JetEnergy}), it is clear to see that AGN BBH mergers occurring at the outer regions of high-mass SMBH are more likely to produce more energetic neutrino bursts that could be possible to be detected by IceCube. Furthermore, the remnant BH mass $M_{\rm BH}$ and the enhancement factor $f_{\rm acc}$ can affect the final detected number of neutrinos, but their influence on the results is not as significant as that of the disk scale height.

{{If the reverse shock becomes a radiation-mediated shock, i.e., $\tau''_{\rm T}\gtrsim1$, Fermi acceleration would be inefficient. As shown in the grey regions of Figure \ref{fig:Number},}} proton acceleration could be prohibited in cases where BBH mergers take place within the radial range of $10^3\lesssim{R}/R_{\rm g}\lesssim10^5$. This corresponds to a region characterized by a density exceeding $\rho\gtrsim10^{-11}-10^{-12}{\rm g}\,{\rm cm}^{-3}$ \citep[see][for details]{Thompson2005}. In such scenarios, the high-density environment results in AGN BBH mergers producing jets with excessively high luminosity and, hence, high jet density. The reverse shock in the jet head could easily become a radiation-mediated shock, given the high internal energy density of the reverse shock being determined by the high jet density, to prevent the Fermi acceleration. When considering the same SMBH mass, AGN BBH mergers with higher masses and larger $f_{\rm acc}$ can generate more energetic jets, which would moderately expand the region where proton acceleration is prohibited. 

{{The observations of the candidate EM counterpart to GW190521 have revealed that the mass of its host SMBH could be $\sim10^8M_\odot$ \citep{Graham2020}.}} Assuming that the jet is directed towards Earth, our simulated results indicate that the number of up-going detected muon neutrinos in IceCube could be as low as a few $10^{-4}$ if GW190521 occurred at the disk migration trap orbit ($\sim100-1000\,R_{\rm g}$) as suggested by \cite{Graham2020}. Consequently, high-energy neutrinos would be challenging to observe. {{Moreover, \cite{Tagawa2023} proposed that the jet breakout signals from GW190521-like BBH mergers, occurring at a radial distance of a few $10^5R_{\rm g}$, could potentially explain the characteristics of the candidate EM counterpart to GW190521.}} At such a radial distance, the number of up-going detected muon neutrinos would be $\gtrsim1$. 

{{By assuming that all AGN BBH mergers are GW190521-like, occurring at $10^5R_{\rm g}$ around a $10^8M_\odot$ SMBH following \cite{Tagawa2023}, we conduct a simplified population simulation to estimate the detection rate of high-energy neutrinos from AGN BBH mergers. In view of that the cosmic evolution of AGN and star formation rate is not significant, we assume that AGN BBH mergers could closely tract the star formation history.}} We randomly generate a population of AGN BBH mergers following $d\dot{N}/dz=\dot{\rho}_0f(z)/(1+z)\times dV(z)/dz$, where $\dot{\rho}_0$ is the local event rate density assuming to be approximately $0.13\,{\rm Gpc}^{-3}\,{\rm yr}^{-1}$ based on GW190521 \citep{Abbott2020GW190521}, $f(z)$ is the normalized star formation rate, and $dV(z)/dz=4\pi D_{\rm L}^2c/H_0(1+z)^2\sqrt{\Omega_\Lambda + \Omega_m(1+z)^3}$ is the comoving volume element. A standard $\Lambda$CDM cosmology with $H_0=67.8\,{\rm km}\,{\rm s}^{-1}{\rm Mpc}^{-1}$, $\Omega_\Lambda = 0.692$, and $\Omega_m = 0.308$ \citep{Planck2016} is applied. {{After considering the beaming effect,}} our simulated detection rate of neutrino events from GW190521-like AGN BBH mergers detected by IceCube is $\sim0.17\,{\rm yr}^{-1}$ ($\sim0.08\,{\rm yr}^{-1}$) when adopting $f_{\rm acc} = 15$ ($f_{\rm acc} = 1$). {Based on the beaming angle, we expect 1 on-axis event per $\sim500$ GW190521-like AGN BBH mergers, implying that we are more likely to achieve multimessenger detection of GWs, EM, and neutrinos during or after the era of GW fifth observing run and IceCube-Gen2.}

\section{Conclusions}

In this {\em Letter}, we present that jets launched by accreting BHs merging in the AGN accretion disks can propagate within the disk atmosphere along with powerful shocks forming at the jet head. Our study suggests TeV-PeV neutrinos can be produced through interactions between protons accelerated by the jet shocks and photons at the jet head via $p\gamma$ interaction. {{Neutrino emissions from internal shock and collimation shock are not taken into account, since the internal shocks of jets from AGN BBH mergers could always be radiation-mediated shocks preventing Fermi acceleration \citep[see Figure 6 in][]{Tagawa2023Highenergy} and collimation shock is uncertain due to the possible presence of cavity or gap around the jets.}} We find that AGN BBH mergers occurring in the outer regions of the disk are more likely to produce more powerful neutrino bursts, whose jet energy can even exceed that of classical long-duration GRBs. These neutrino bursts could be detected by IceCube if the highly-collimated jets are on-axis and neutrinos enter the IceCube instrument from the up-going direction.

Using the host AGN properties of the potential GW190521 EM counterpart as an example, we expect $\gtrsim1$ up-going neutrinos detectable by IceCube for on-axis GW190521-like BBH mergers occurring at $R\gtrsim10^5R_{\rm g}$. Events occurring at such a distance range may also potentially explain the observational results of the GW190521-associated EM counterpart \citep[e.g.,][]{Tagawa2023}. {Neutrino bursts from GW190521-like AGN BBH mergers could be rare with a simulated detection rate $\lesssim0.2\,{\rm yr}^{-1}$ because of the low event rate density and highly-collimated jets.} {{The disk model we used, as presented in \cite{Thompson2005}, suggested that disks can be marginally stable and avoid fragmentation over a wide range of radii. Because the transition from unstable galactic disks to accretion disks is still unclear, e.g., \cite{Haiman2009} suggested that disks at large radii of a few $10^4\,R_{\rm g}$, considering the classical \cite{Shakura1976} model, may not be stable. Thus, the observations of high-fluence neutrinos from AGN BBH mergers in such cases can only be achievable if mergers occur at distances ranging from a few $10\,{\rm Mpc}$ to a few $100\,{\rm Mpc}$.}}

We predict that neutrino bursts from AGN BBH mergers could be detected by IceCube following the observation of BBH GWs. 
{{The detections of high-energy neutrino signals would occur before the EM breakout signals.}} {{Thus, the recent neutrino events, which triggered before GWs, reported by IceCube in O4 \citep[e.g.,][]{2023GCN1,2023GCN2,2023GCN3} can not be originated from AGN BBH mergers.}}

\section*{Acknowledgements}

{JPZ thanks an anonymous referee for thoughtful and constructive suggestions.} JPZ thanks Todd A Thompson, Ilya Mandel, Bing Zhang, Di Xiao, Yun-Wei Yu, Yacheng Kang, and Di Zhang for helpful comments.

\section*{Data Availability}

The data generated in this work will be shared upon reasonable request to the corresponding author.

\appendix

\section*{Schematic Picture}

\begin{figure*}
     \centering
     \includegraphics[width=1\linewidth]{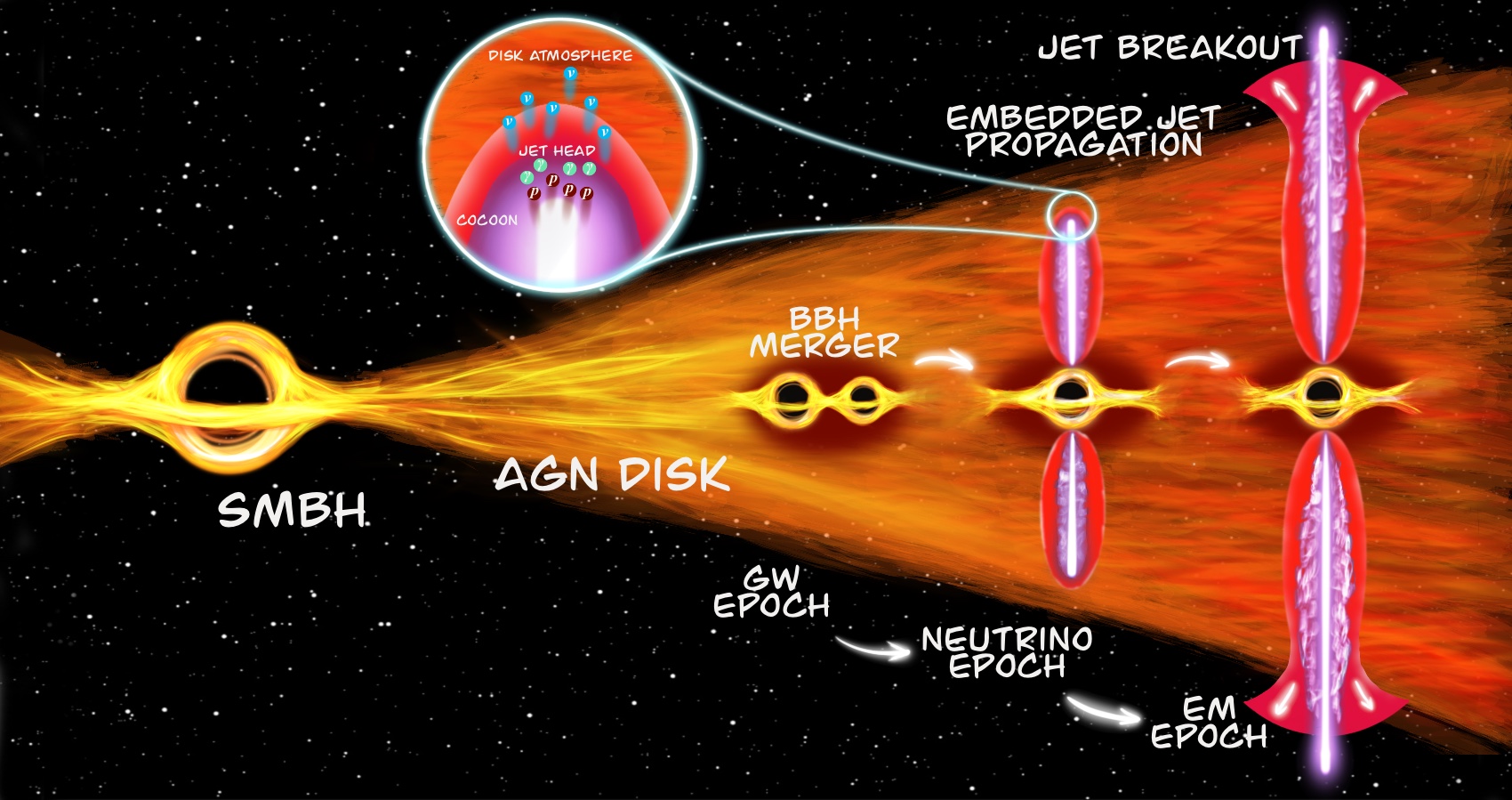}
     \caption{Schematic picture of the physical processes of jet propagation and neutrino production for AGN BBH mergers. The crimson, green, and blue particles in the enlarged view represent protons, photons, and neutrinos, respectively.}
     \label{fig:Schematic}
\end{figure*}

Figure \ref{fig:Schematic} illustrates the physical processes of jet propagation and neutrino production for AGN BBH mergers. {{According to \cite{Wang2021BBH}, powerful Blandford–Znajek jets can be generated directly after the BBH merger within the AGN disk if the viscosity effect rapidly removes the angular momentum of the binary system when they merge. \cite{Bartos2017,Wang2021,Kimura2021,Tagawa2022,Chen2023} suggested that the accretion of BBH binaries can power outflows or jets to potentially create outflow cavities or open gaps around the binaries in AGN disks. Even in the presence of a gap or cavity around the BBH merger, the jets can be launched and still interact with the disk materials as the remnant BHs traverse the disk \citep{RodrguezRamirez2023} or after re-orientations of jets by the mergers \citep{Tagawa2023}.}} Powerful neutrino bursts from AGN BBH mergers would be always occurred after mergers. {{Then, after the breakout of the jets from AGN BBH mergers, photons at the jet head can escape and, hence, neutrino productions would cease. Thus, neutrino bursts can be the}} precusor signals before the EM breakout signals.

\bibliographystyle{mnras}
\bibliography{ms} 

%\appendix



\bsp	
\label{lastpage}
\end{document}